# Machine learning based biomedical image processing for echocardiographic images

Ayesha Heena[1] · Nagashettappa Biradar[1] · Najmuddin M. Maroof[2] · Surbhi Bhatia[3] · Rashmi Agarwal[4] · Kanta Prasad[5]

**Abstract**
The popularity of Artificial intelligence and machine learning have prompted researchers to use it in the recent researches. The proposed method uses K-Nearest Neighbor (KNN) algorithm for segmentation of medical images, extracting of image features for analysis by classifying the data based on the neural networks. Classification of the images in medical imaging is very important, KNN is one suitable algorithm which is simple, conceptual and computational, which provides very good accuracy in results. KNN algorithm is a unique user-friendly approach with wide range of applications in machine learning algorithms which are majorly used for the various image processing applications including classification, segmentation and regression issues of the image processing. The proposed system uses gray level co-occurrence matrix features. The trained neural network has been tested successfully on a group of echocardiographic images, errors were compared using regression plot. The results of the algorithm are tested using various quantitative as well as qualitative metrics and proven to exhibit better performance in terms of both quantitative and qualitative metrics in terms of current state -of- the-art methods in the related area. To compare the performance of trained neural network the regression analysis performed showed a good correlation.

## 1 Introduction

Echocardiography is an important diagnostic procedure for the cardiologist to evaluate the structure and function of heart. Being non-invasive, low in cost and use of non-ionizing radiations, the 2D echocardiography has been largely applied in evaluation of cardiac

✉ Ayesha Heena
ayeshaheena31@gmail.com

Extended author information available on the last page of the article

Springer

functions. In this study a machine learning algorithm employing neural network is trained to identify the texture and shape features of echocardiographic images, based on preliminary knowledge and can be used to trace the boundary of the region of interest. To test the accuracy of the algorithm regression analysis is done and the regression plot showed good correlation. The speckle noise is the main reasons that corrupt the biomedical images when they are captured. A neural network consists of a large number of processing elements called neurons. Each neuron has an internal state called activation or activity level, which is a function of the inputs it has received. Every neural network has knowledge which is contained in the values of the connection weights. Modifying the knowledge stored in the network as a function of experience implies a learning rule for changing the values of the weights. Typically, a neuron sends its activation as a signal to several other neurons. A neuron can send only one signal at a time, although that signal may be broadcast to several other neurons

$$\text{Medical imaging} + \text{Advanced Machine learning} = \text{Transforming patient care} \quad (1)$$

This equation above explains us how the importance of Machine learning in health care has evolved in past few years and how it transformed the patient care systems. There is a bit shortage of Radiologist in India, there is a need of Machine learning algorithms that can imitate Radiologist. Machine learning more specifically the field of predictive modelling is primarily concerned with minimizing error of a model, or making the most accurate predictions possible at the expense of explain ability. Machine learning algorithm is used to solve both classification and regression problems. It is easy to implement and understand.

### 1.1 Motivation and contributions of the work

Manual process of segmentation and border detection is in general subjective, time consuming and unreliable process. Many techniques have been developed to automate or semi- automate the processes. The segmentation process may be manual, semi-automatic or automatic depending on the extent of operator intervention. For automatic segmentation process image analysis is usually based on 'a priori' knowledge of the structures [1]. Since most of the images in echocardiography have poor quality. For automatic process one method is use of neural networks. The features of the texture smooth area might be attenuated significantly and also its differential results might be very near to zero.

Most of the segmentation techniques linearly attenuates the features of the texture and this will also fail to preserve them in the given area. The need to preserve the textural features in the smooth areas is very essential in biomedical image segmentation (all biomedical images have intensity inhomogeneity) missing of even minute details could lead to wrong analysis and diagnosis which is very dangerous for patients and risky for life of the patients.

The following are the main contributions of the work carried out in this article:

- In the proposed work first the input echo image is denoised so as to reduce the speckle noise which has multiplicative impact on image hence denoising is an inevitable step in image processing of biomedical images.
- KNN based segmentation is the next part of the algorithm after pre-processing, feature extraction and classification of the pixels is done. Results shows we could mark the boundaries in the image. Features extracted are gray level co-occurrence matrix features.
- Neural network pixel classifier is used to classify the pixels into inter or intra pixels.

The rest of the paper has been organized as follows, Section 2 describes the related work, Section 3 gives the methodology used in the article and Section 4 highlights the results and discussion and Section 5 concludes the paper with future directions and open challenges.

## 2 Related work

This section presents the previous studies carried out related to KNN algorithm. In [2, 16] the authors presented how to choose optimal value of k for classification of medical images. In [4] the work discusses on neighborhood criteria and the component analysis of the neighborhood. Feature selection models to resolve high dimensional classification of data is discussed in [3]. The authors in [5] have proposed a differentially nearest neighbor classification and with the purpose of sensitive data to protect privacy. The analysis includes sensitivity analysis carried out on different image datasets. The reference [14] a study on distance functions is carried out to understand the impact on KNN Classification for medical datasets. Various distance function metrics such as Euclidean, cosine, Chi-square and Minkowski were used. The study concluded Chi-square proves to be the best distance measure. Prediction error is inevitable and unable to resolve sparsity and imbalance in classes using KNN classifiers. The consideration is made that in the smooth area of the image there will be no significant changes in the gray scale.

For the measure of fractional calculus, the frequently used common definitions are R_L, Grünwald-Letnikov (G-L) and Caputo [11] definitions. The Euclidean distance measure is very frequently used and commonly applied definitions for the fractional calculus. The Euclidean distance measure in the definition of G-L function. First, we need to understand the type of data in our data set. Then explore the data to understand the nature and quality and explore the relationship amongst the data elements i.e., relationship between features. Apply the preprocessing as needed.

Once the data is prepared for modelling, (Structured representation of raw input data into a meaningful pattern is called a model) then learning starts. The process of assigning model and fitting a specific model to the data set is called training of the model. First the input data is divided into training data and test data, train the model based on training data and apply to test data. Performance of the model is then evaluated. Based on options available, specific action is taken to improve the performance of the model. A machine learning algorithm creates its cognitive capability by building a mathematical formulation or a function based on features in the input data set [10–12]. Models for supervised learning are predictive models, try to predict certain value using values in the input data set. The learning model attempts to establish a relation between feature being predicted and the predictor features. Predictive models in turn predict the value of a category or class to which data instance belongs to. The models which are used for prediction of target features of categorical value are known as classification models. The target feature is known as a class and the categories to which classes are divided are called levels. The model used in research is K Nearest Neighbor (KNN) method.

### 2.1 Research gaps

Survey shows that various algorithms developed for segmentation and classification with feature extraction suffers from artifacts, speckle noise, improper marking of boundary fail preserving edges etc.

KNN is a simple supervised machine learning algorithm. In KNN some prior data called the training data are given which classifies coordinates into groups based on features extracted. Machine learning practitioners spend significant amount of time in different feature engineering activities. Selecting right features has a critical role to play in the success of machine learning model.

Existing algorithms shows poor performance in terms of the MSE, PSNR, SSIM analysis metrics. Overcoming the artifacts and shadowing effects inherently present in the echo images is a huge challenge. If we could overcome this challenge then better metrics is obtained in terms of MSE, PSNR and SSIM.

## 3 Methodology

K-Nearest neighbor (KNN) is one of the most essential classification algorithms used in the research, the algorithm selects K nearest training samples for a test sample and then predicts the test sample with the major class among K nearest training samples [18]. We can choose the right K for our data by trying several K's and picking the one that works best. The regression plots obtained in linear regression analysis estimates the coefficients of the linear equations which involves one or more independent variables, where the variable which you want to predict is known as dependent variable and variable you are using to predict the other variables is called the independent variable. The simple regression is a model which has a single regressor x which has a relationship with response y that is a straight line. Classification task depends upon labelled datasets that is humans must transfer their knowledge to the datasets in order for a neural network to learn the correlation between labels and data [15]. This is known as supervised learning. Any labels that humans can generate outcomes that you care about and which correlate to data can be used to train the neural networks. K-nearest neighbors on the other hand, is efficient and can be implemented easily as per the user needs, [17] but the limiting factor of this algorithm is the computational speed is very less when the number of data sets increases, which also interprets the computational speed, is quite high for the fewest amount of data for the processing. This KNN algorithm operates with the basic principle of the amount of weights or the calculated distance among the query and the selective examples in the data set by the specific number which is variable defined as 'k', and this 'k' is the closest among the query, [9, 19, 20] then the selection will be done on the relevant frequency label of the data for the classification of pixel in the image data or it identifies the average labels for the regression operations, [21, 22] in the regression based operation it can be quoted that the selection of the suitable 'k' for the required data serve for the best of the 'k' states which will be the best one for the exchange and selection of pixel data [6, 7]. Whereas the KNN algorithm is the best for the less amount of the data pixels for the query and processing [8, 13]. Weighted directed graph is a perfect term to define KNN graph. The algorithm KNN is based on the distance function voting. Categorical values can be converted into numerical values and vice versa. If the training data set is small low variance models perform better to avoid model overfitting. When training data is large low bias models are preferred. Few models can be used for both classification as well as regression. The subset in the input data is used as test data for evaluating the performance of trained model. The test data is used only for once, after the model is refined and finalized, to measure and report the final performance of the model as reference for future learning.

To evaluate the performance of the model in case of supervised learning which is used in research, the number of correct classifications or predictions made by the model need to be recorded. Based on which accuracy is calculated for the model. Also, the regression model which ensures the difference between predicted and actual values is low is considered as good model. We can obtain the higher and superior image denoising capabilities, by using the fractional differential-based operators rather than the integer differential-based operators.

### 3.1 Image Denoising

As image is a two-dimensional representation of any data given as (x, y) which will be sensitive to the various kinds of noises which are destructive in nature which corrupts the input image. In our point of interest as this research is oriented to the ultra sound images which are a biomedical input data. But as a major challenge these biomedical data i.e., ultrasound images are highly affected by the multiplicative noise and the corrupted ultrasound image can be represented as

$$f(x,y) = g(x,y) \cdot n(x,y) \qquad (2)$$

Where,

$f(x, y) = >$ corrupted noisy ultrasound image.
$g(x, y) = >$ image prior to the affecting of multiplicative noise.
$n(x, y) = >$ multiplicative noise i. e. , speckle noise.

The ultrasound image with the spatial coordinates (x, y) are restricted to the M x N dimensions which will be of the same dimension for the raw image prior affecting of noise and after corrupting the input data. The multiplicative noise (speckle noise) also proportional with the data following the property of two-dimensional structure. Hence it is highly important to cut-down the impact of the multiplicative noise as a preprocessing step on the ultrasound image by employing the available mathematical operations on it, one of the suitable operations will be the logarithmic operator.

This is represented as

$$\log[f(x,y)] = \log[g(x,y) \cdot n(x,y)] \qquad (3)$$

with the basic property of the logarithms the equation can be rewritten and applied as

$$\log[f(x,y)] = \log[g(x,y)] + \log[n(x,y)] \qquad (4)$$

The crucial advantage of applying the logarithmic operator on the affected noise is to cut-down the effect of noise on the image by the similar operation of decoupling for the minimization of impact of multiplicative noise, hence it will result in the separation of noisy residue from the image under consideration there by the noise free biomedical image can be retrieved for further operation [10].

### 3.2 Image data

In this paper standard clinical database from Medical Information Mart for Intensive Care III (MIMIC-III) is a large, freely-available database comprising deidentified health-related data associated with over 40,000 patients who stayed in critical care units of the Beth Israel Deaconess Medical Center between 2001 and 2012 is used. The MIMIC-III Clinical Database is available on Physio Net (https://doi.org/10.13026/C2XW26). Though deidentified, MIMIC-III contains detailed information regarding the care of real patients, and as such requires credentialing before access. To allow researchers to ascertain whether the database is suitable for their work, they have manually curated a demo subset, which contains information for 100 patients also present in the MIMIC-III Clinical Database. These patients were selected randomly from the subset of patients in the dataset who eventually die. This project was approved by the Institutional Review Boards of Beth Israel Deaconess Medical Center (Boston, MA) and the Massachusetts Institute of Technology (Cambridge, MA). MIMIC-III is a relational database consisting of 26 tables. The MIMIC-III demo provides researchers with an opportunity to review the structure and content of MIMIC-III before deciding whether or not to carry out an analysis on the full dataset. The data files are distributed in comma separated value (CSV) format following the RFC 4180 standard. Further for analysis of the results obtained for the standard database, it is also tested and applied for the other standard database of the image processing i.e., six different images as described in the results and discussion in next section in Fig. 3.10.

### 3.2.1 Features

Very important characteristic that is used to identify the objects for any region is the texture. Based on the spatial dependencies, the texture of the image is characterized by major components such as gray level co-occurrence Matrix features for the biomedical image that is the Contrast, Homogeneity (H), entropy & Local Homogeneity (LH).

If I is the grayscale image, the position of each pixel is given by $s \equiv (x, y)$ in I and $t = (\Delta x, \Delta y)$. The co-occurrence matrix M is m x m matrix. The pair of gray levels (i,j) is

$$M(i, j) = \text{card}\{(s, s+t) \in R^2 \mid I(s) = I, I(s+t) = j\} \quad (5)$$

M➔ concurrent matrix.
Card➔ predefined MATLAB function for the data exchange.
S➔ position of the pixel.
I➔ image data under consideration.
t➔ translation vector.
i,j ➔ gray level.

The Contrast, when the scale of local texture is larger the distance

$$C = \sum_{k=0}^{m-1} k^2 \sum_{|i-j|=k} M(i,j) \quad (6)$$

k➔ neighboring pixels.

Homogeneity is represented as

$$H = \sum_i \sum_j (M(i,j))^2 \quad (7)$$

Entropy is the measure of randomness: close to either 0 or 1

$$E = \sum_i \sum_j (M(i,j)) \log(M(i,j)) \quad (8)$$

Local Homogeneity (LH) is given as

$$LH = \sum_i \sum_j \frac{M(i,j)}{1+(i-j)^2} \quad (9)$$

The above mentioned four various features/parameters of the biomedical image is very essential for the KNN based image segmentation which will be done at the pixel level for the query and reference pixel, in this regard the first parameter contrast given by the Eq. (6) in which the texture decision basically done on the contrast of test image with the neighboring pixels. The secondary parameter is homogeneity in which it decides the pixel region and the background image which is not concern, on the other hand it is most required for the data extraction from the region of interest of the biomedical image data which is given by the Eq. (7). Next parameter is the entropy which is given by the eq. (8) in which it plays a decision parameter for the 1 or 0 for pixel exchange. Last parameter Local Homogeneity (LH) in which similar to the homogeneity which is the done among the inter and intra pixel of analysis given by eq. (9).

The KNN based approach for the segmentation can be listed into four different phases namely.

Pre-Processing
Feature extraction
Classification
Post processing

*Phase 1: Pre-processing*
The removal of noise and the conversion of the image to grayscale is completed in the process of preprocessing. Adaptive fractional order integral filter is used for image denoising as discussed in [10].

*Phase 2: Feature Extraction*
The main intention of feature extraction is to reduce the original set of data base that are distinguishable on the patterns of input. The texture feature extraction method is used to extract the features such as correlation, Homogeneous, entropy, contrast and local Homogeneous.

*Phase 3: classification*
The pixel is classified into the similar class by the training data which has the intensity of same level this classification is done in nearest neighborhood algorithm. To avoid the errors that occurs because of the single neighborhood outliers of other class, and improve the robustness of the approach, the classifier works with k patterns and hence it is called K nearest neighborhood algorithm.

*Phase 4: Post Processing*
The extraction of abnormal region from the image is done in the step of post processing. The removal of the cult involves the first step, it involves removal of grey matter and hence the abnormal region is extracted.

### 3.2.2 Neural networks approach

The main reasons for the extensive usage of Neural Networks (NN's) are their powerful aspects and ease of use. A simple approach is to firstly extract a feature set representing the biomedical image details (like noise, pixel value), with several samples from different database.

The second step is for the NN to learn the relationship between an input image and its class among other regions. Once this relationship has been learned, the network can be presented with test input that can be classified as belonging to a particular pixel. NNs therefore are highly suited to modeling global aspects of input data.

In the classifier approach, with respect to neural network-based approach it is found that the number of input layers were kept as 108 and hidden layers as 39 and at the output.

## 4 Results and discussions

The proposed method for the biomedical image processing with the various operations can be summarized as shown in Fig. 1, the analysis of the complete process consists of the few steps which are given by:

1. Consider any biomedical image or standard image as the input function for the proposed system.
2. Image segmentation and Selection of R-L algorithm for the preprocessing of the input image is done so as to denoise it.
3. Overlapping the function (x, y) with the fractional mask either 3 X 3 or 5 X 5.
4. Feature extraction with the membership function for the image region from each pixel and employing the KNN method of pixel segmentation.
5. Applying the filter for each pixel and to obtain the decision it is fed to the Neural Networks for the inter and intra pixel data classifier.
6. Repeat the steps 2–5 among all the input data.
7. Calculation of PSNR, MSE, variance and SSIM from the input image.

Figures 2 and 3 shows the sample output for the Echo image which is selected from the standard database which provides us the dashboard showing the respective values of PSNR and MSE for the particular image in which the outputs are divided into three different frames showing the various stages of the input processing as raw input image, noisy image, fuzzy processed image, other two windows showing the MSE, PSNR and SSIM for the particular window-based performance image. The PSNR obtained is high, MSE is very low and SSIM is very good in comparison to other methods. Which also proves that the processing resulted with significant removal of noise and artifacts.

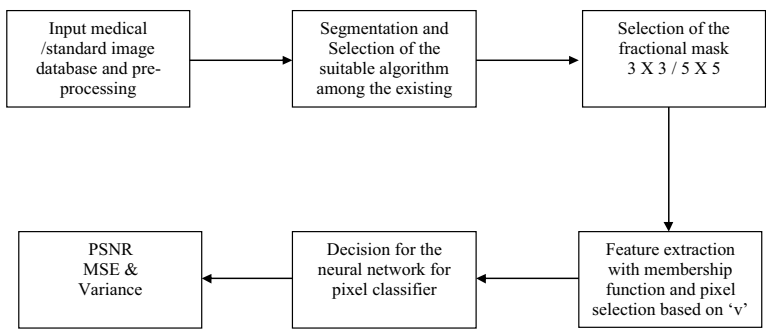

**Fig. 1** Proposed System for biomedical image processing, feature extraction and segmentation

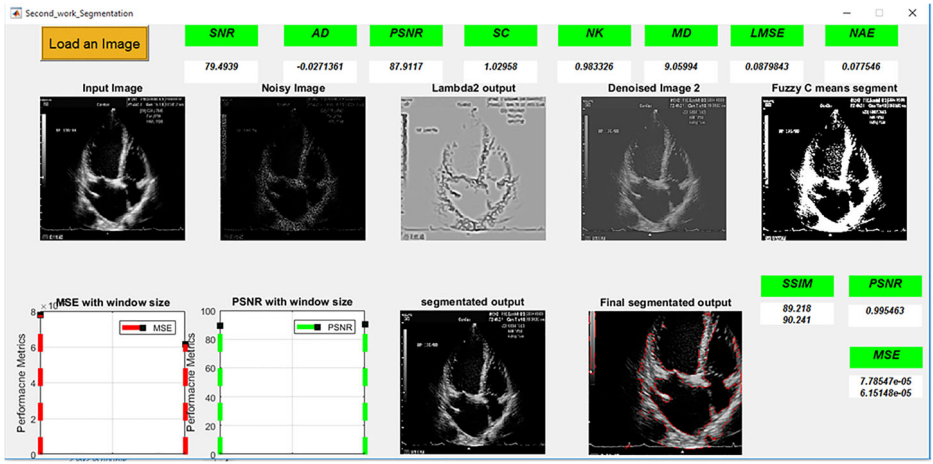

**Fig. 2** Echo image analysis with detailed approach of image processing and calculation of the PSNR & MSE

As per the proposed algorithm, at the classfier stage it is proposed to implement neural network as the prefered classifier its representation is as shown in Figs. 4 and 5, in which it is provided with the details of training and testing phase with the training performanance.

In the proposed scheme for the biomedical image processing which are discussed in Figs. 6 and 7, in which the biomedical image segmentation is done using the KNN algorithm the original image and segmented image can be represented in Figs. 6 and 7, Fig. 6(a) shows the original image from database under consideration after identifying from the background noise subtraction by applying the filtering technique, Fig. 6(b) shows the fully segmented biomedical image from the database by applying KNN algorithm. The comparative graph for the LMSE & Variance for the various images

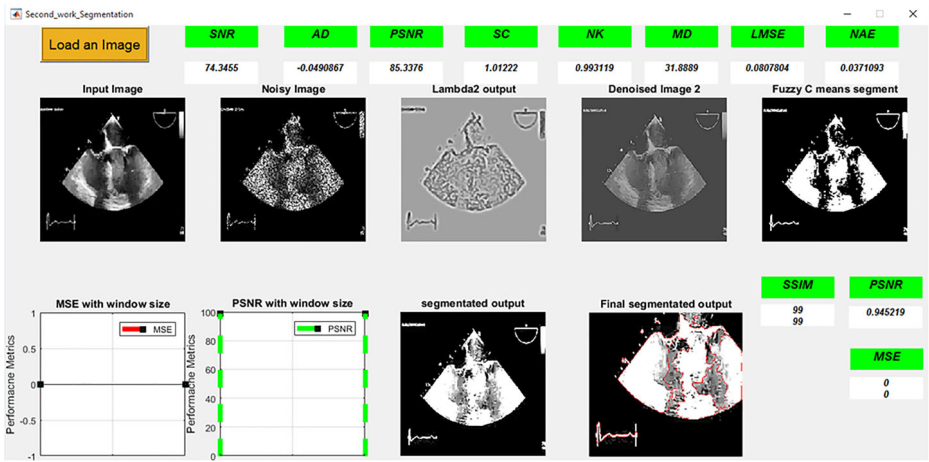

**Fig. 3** Echo image analysis with detailed approach of image segmentation and calculation of the PSNR & MSE

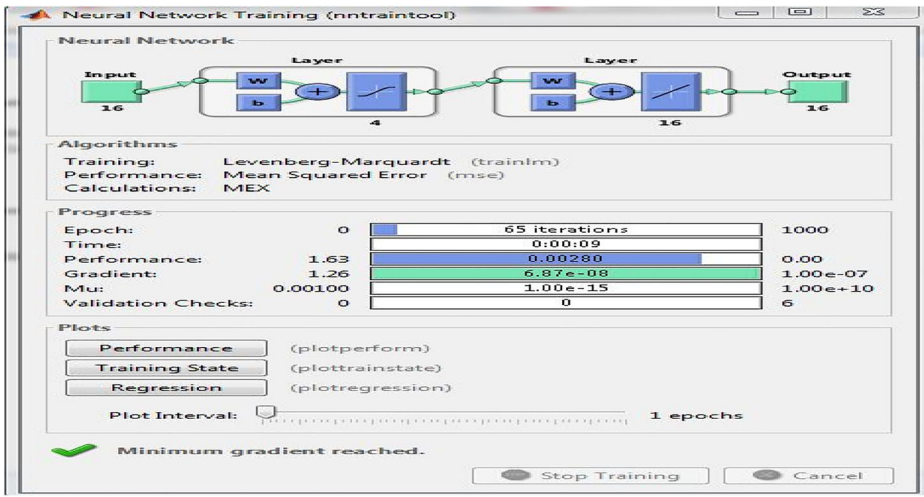

Fig. 4 sample output of the neural network as classifier operation from the database collection is as shown in Fig. 7. For the standard image database for the various analysis upon the steps described by the calculation of SNR & PSNR is as shown in Fig. 8.

The desired algorithm is tested with the non-medical image database also, and its comparative analysis is as done with the existing and proposed algorithms as shown in Fig. 9 and its tabulated values are as shown in Tables 1 and 2. For analysis it is considered only with three test input images (synthetic namely Camera man, Checkerboard) and Ultrasound (US) image. With the developed algorithm the performance is

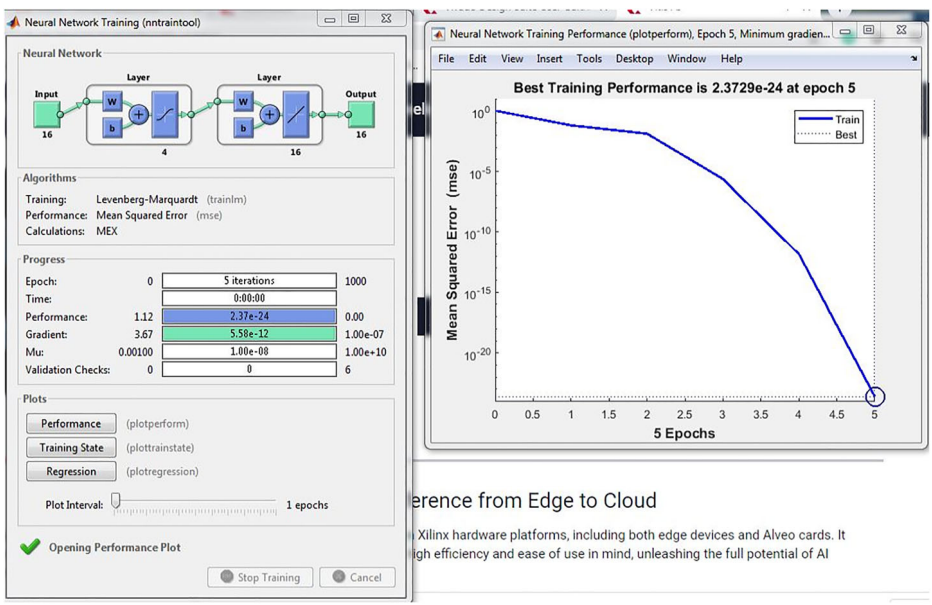

Fig. 5 Neural Network with best training performance

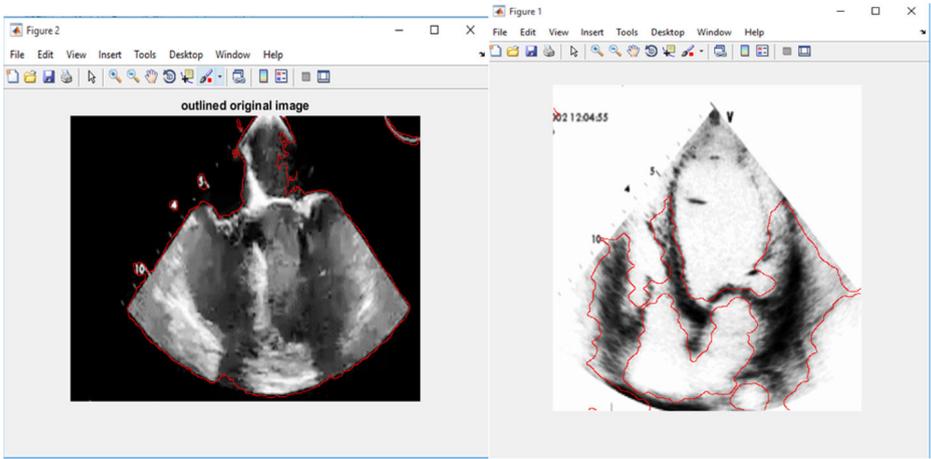

(a)                    (b)

**Fig. 6** **a** original image with contour marked. (**b**) segmented image with the KNN

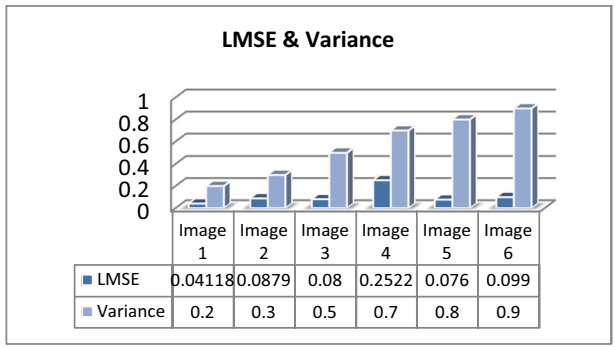

**Fig. 7** comparative graph of LMSE & Variance

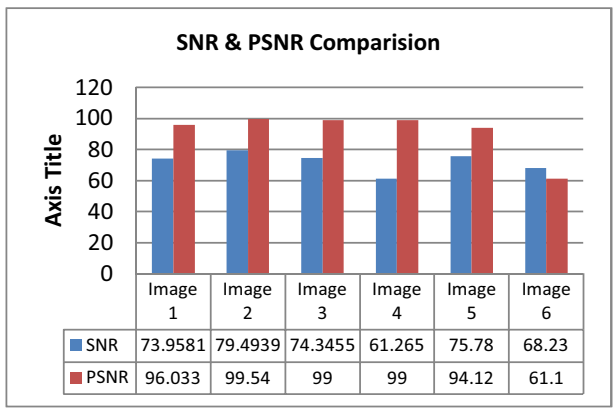

**Fig. 8** Comparative graph of SNR & PSNR

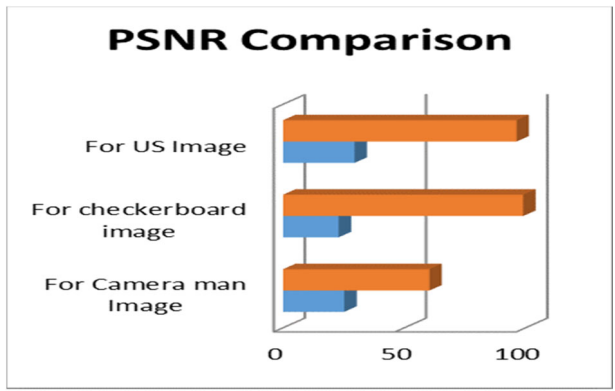

(a)

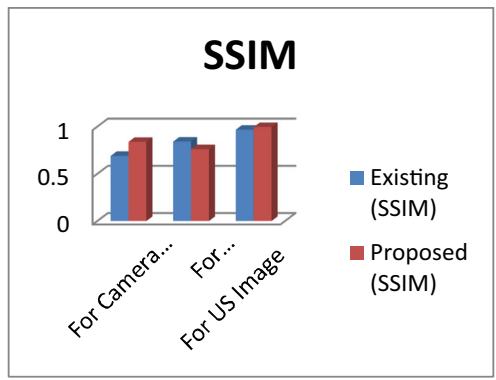

(b)

Fig. 9 comparative graph for ordinary images (a) PSNR (b) SSIM

measured using the Structural Similarity Index Matrices (SSIM), this similarity index is used as the similarity tool for the measurement of the performance and efficiency, it has to be noted that there exists a remarkable enhancement of the similarity index among the standard database. Figure 10 gives a summary of results obtained for the set of echo

Table 1 Comparative analysis of standard database for normal images comprising of values corresponding to parameters

| Parameters | MSE | | PSNR | | SSIM | |
|---|---|---|---|---|---|---|
| Input Image name | Existing | Proposed | Existing | Proposed | Existing | Proposed |
| Camera man Image | 0.193 | 0.198 | 25.26 | 60.339 | 0.69 | 0.8383 |
| Checkerboard image | 0.0033 | 7.7854 e-05 | 22.94 | 99 | 0.8418 | 0.7619 |
| US Image | 0.00487 | 7.7855 e-05 | 29.5137 | 96.26 | 0.9672 | 0.9954 |

Table 2 Summary of parameters of calculation for six medical images from standard database

| Database | MSE | | PSNR | | SSIM | |
|---|---|---|---|---|---|---|
| | Existing | Proposed | Existing | Proposed | Existing | Proposed |
| Image1 | 0.193 [6] | 0.04118 | 37.04 [6] | 96.033 | 0.967 [6] | 0.979581 |
| Image2 | 0.368 [10] | 0.0879 | 28.27 [10] | 99.54 | 0.641 [10] | 0.794939 |
| Image3 | 0.00487 | 0.080 | 29.5137 | 99 | 0.7672 | 0.743455 |
| Image4 | 0.67 | 0.2522 | 76.3 | 99 | 0.542 | 0.61265 |
| Image5 | 0.45 | 0.076 | 67.8 | 94.12 | 0.63 | 0.7578 |
| Image6 | 0.63 | 0.099 | 81.2 | 61.10 | 0.574 | 0.6823 |
| Average | 0.32 | 0.10 | 50.5 | 91.2 | 0.641 | 0.723 |

images and parameter evaluation. Table 3 gives the details of features extracted using KNN algorithm in terms of contrast, homogeneity, Entropy and Local homogeneity. Table 4 gives the overall performance of the algorithm in terms of Accuracy, Sensitivity

| Database Image | Input image | Preprocessed output Image | Segmented output Image | Parameters |
|---|---|---|---|---|
| Image1 | 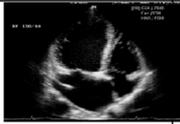 | 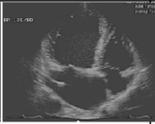 | 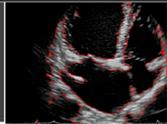 | SNR :79.49<br>PNSR:87.91<br>LMSE:0.087<br>SSIM:90.24<br>Variance:0.2 |
| Image:2 | 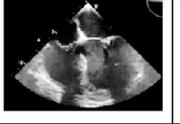 | 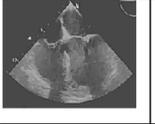 | 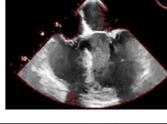 | SNR :74.34<br>PNSR:85.33<br>LMSE:0.080<br>SSIM:99<br>Variance:0.3 |
| Image:3 | 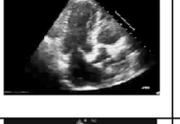 | 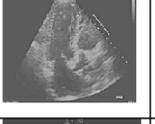 | 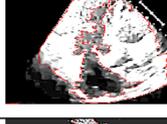 | SNR :75.78<br>PNSR:86.05<br>LMSE:0.076<br>SSIM:89.2<br>Variance:0.5 |
| Image:4 | 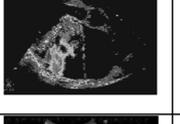 | 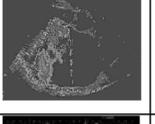 | 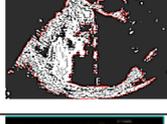 | SNR :73.95<br>PNSR:85.14<br>LMSE:0.24<br>SSIM:61.98<br>Variance:0.7 |
| Image:5 | 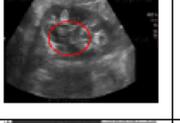 | 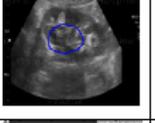 | 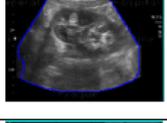 | SNR :93.55<br>PNSR:89.24<br>LMSE:0.034<br>SSIM:91.65<br>Variance:0.8 |
| Image:6 | 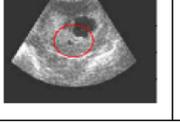 | 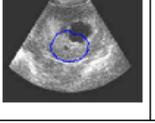 | 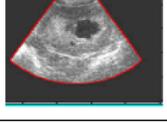 | SNR :93.95<br>PNSR:89.34<br>LMSE:0.054<br>SSIM:81.78<br>Variance:0.9 |

Fig. 10 Summary of results of segmentation and parameter evaluation

Table 3  Features extracted for KNN based segmentation

| S.NO | Contrast | Homogeneity | Entropy | LH |
|---|---|---|---|---|
| 1 | 0.978 | 0.897 | 0.242 | 0.89 |
| 2 | 0.965 | 0.824 | 0.424 | 0.86 |
| 3 | 0.925 | 0.923 | 0.484 | 0.89 |
| 4 | 0.956 | 0.845 | 0.399 | 0.88 |
| 5 | 0.922 | 0.822 | 0.459 | 0.85 |

Table 4  Results of performance of algorithm (average values)

| S.No. | Images | Accuracy (%) | Sensitivity (%) | Specificity (%) |
|---|---|---|---|---|
| 1. | Synthetic | 89 | 97 | 25 |
| 2. | Echo | 88 | 63 | 81 |

and specificity (in percentage) average values for tested synthetic images as well as echo images.

# 5 Conclusion

The algorithm presented in the paper used the pixels of a region that resulted on entries to neural net. This procedure came out satisfactorily proving that it is possible to train the net to recognize the image models.

Finally, algorithm is demonstrated with neural network for the classification of the image and the comparative analysis of Mean Square Error, PSNR, SSIM and variance from experimental results and analysis has proven that the algorithm is better than the existing various methods for standard medical image database and other image databases. It has to be noted that the improvement of the proposed algorithm with addition of the neural network as a classifier, which played an important role.

It should also be noted that developed algorithm can also be applied to any of the real time or other untrained database for the applications.

## 5.1 Future research directions and open challenges

As a future enhancement an advanced algorithm can be designed for the video applications, satellite images and other format of images, even as next improvement it can also extended to the comparative analysis among another available database.

Further it is desired to perform abnormality classification of echo images into mild, moderate, and severe abnormality based on features extracted using ANN. Few more features are to be added and train the network to get high accuracy results of abnormality classification.

In future we could explore many more applications of machine learning by taking into account the energy features and applying classifiers using NN to detect the abnormality, also classify it in terms of whether the abnormality is mild, moderate or critical. If this is possible it would a boon for Doctors/Cardiologists/Radiologists as an assessment in their diagnosis.

**Acknowledgments** The author would like to thanks my family for providing the constant support for the preparation of this article. The author would like to extend the deepest and sincere thanks to BKIT, VTU and KBNU.

## Declarations

**Competing interests** The authors declare that they have no conflicts of interest.

## Affiliations

Ayesha Heena[1]  · Nagashettappa Biradar[1] · Najmuddin M. Maroof[2] · Surbhi Bhatia[3] · Rashmi Agarwal[4] · Kanta Prasad[5]

Nagashettappa Biradar
nmbiradar@gmail.com

Najmuddin M. Maroof
ecemaroof99@gmail.com

Surbhi Bhatia
surbhibhatia1988@yahoo.com

Rashmi Agarwal
drrashmiagrawal78@gmail.com

Kanta Prasad
tokpsharma@gmail.com

[1] Department of Electronics and Communication, BKIT Bhalki Karnataka/VTU Belagavi, Karnataka, India

[2] Department of Electronics and Communication, KBN College of Engineering Kalaburagi Karnataka/VTU Belagavi, Karnataka, India

[3] Department of Information Systems, College of Computer Science and Information Technology, King Faisal University, Al Hasa, Saudi Arabia

[4] Department of Computer Application, Manav Rachna International Institute of Research and Studies, Faridabad, India

[5] Department of Computer Science, GL Bajaj Group of Institutions Mathura, Mathura, India